\documentclass[showpacs,twocolumn,showkeys,amsmath,amssymb,pra,nofootinbib,subscriptaddress,longbibliography]{revtex4-1}

\usepackage{amsmath,amssymb,amsfonts}
\usepackage{graphicx}
\usepackage{txfonts}
\usepackage{epstopdf}
\usepackage[T1]{fontenc}
\usepackage[unicode]{hyperref}
\hypersetup{
   unicode=true,          
   a4paper=true,
   plainpages=false,
   pdftitle={Title of PDF},    
   pdfauthor={Author of PDF},     
   pdfsubject={Subject of PDF},   
   colorlinks=true,       
   citecolor=blue,        
}
\begin{document}

\title{Progress towards quantum-enhanced interferometry with harmonically trapped
  quantum matter-wave bright solitons
}

\author{Bettina Gertjerenken}
\email{b.gertjerenken@uni-oldenburg.de}
\affiliation{Department of Mathematics and Statistics, University of Massachusetts Amherst, Amherst, MA 01003-9305, USA}
\affiliation{Institut f\"ur Physik, Carl von Ossietzky Universit\"at, D-26111 Oldenburg, Germany}

\author{Timothy P.\ Wiles}\thanks{Now at Peratech 
  \url{http://www.peratech.com/}}

\affiliation{Joint Quantum Centre (JQC) Durham--Newcastle, Department of Physics, Durham University, Durham DH1 3LE, United Kingdom}

\author{Christoph Weiss}
\email{christoph.weiss@durham.ac.uk}
\affiliation{Joint Quantum Centre (JQC) Durham--Newcastle, Department of Physics, Durham University, Durham DH1 3LE, United Kingdom}

\date{\today}

 \begin{abstract}
We model the dynamics of attractively interacting ultracold
bosonic atoms in a quasi-one-dimensional wave-guide with additional
harmonic trapping. Initially, we prepare the system in its ground
state and then
shift the zero of the harmonic trap and switch on an additional narrow
scattering potential near the center of the trap. After colliding with
the barrier twice, we propose to measure the number of atoms opposite to the initial
condition. Quantum-enhanced
interferometry with quantum bright solitons allows us to predict
detection of an offset of the scattering
potential with considerably increased precision as compared to
single-particle experiments. In a future experimental realization  
this might lead to measurement of weak forces caused, for example,
by small horizontal gradients in the gravitational potential --- with
a resolution of several micrometers given essentially by the size of
the solitons. 
Our numerical simulations are based on the rigorously proved effective
potential approach developed in [Phys.\ Rev.\ Lett.\ \textbf{102}, 010403
(2009) and Phys.\ Rev.\ Lett.\ \textbf{103}, 210402 (2009)]. We choose
our parameters such that the prerequisite of the proof (that the
solitons cannot break apart, for energetic reasons) is always
fulfilled, thus exploring a
parameter regime inaccessible to the mean-field description via the
Gross-Pitaevskii equation due to Schr\"odinger-cat states occurring
in the many-particle quantum dynamics.
\end{abstract}


\maketitle


\section{Introduction}

Matter-wave bright solitons can be used for
interferometry by profiting from their mean-field
description based on the Gross-Pitaevskii equation (GPE) --- primarily
by the fact that attractive interactions
prevent the wave-packets from spreading
\cite{MartinRuostekoski2012,PoloAhufinger2013,McDonaldEtAl2014,HelmEtAl2015,SakaguchiMalomed2016}.  Matter-wave
bright solitons\footnote{Our solitons are not
  solitons in the strict sense but rather solitary waves. However,
  these solitary waves can behave very similar to solitons~\cite{MartinEtAl2008b}.} are
investigated experimentally both for attractive
interaction~\cite{KhaykovichEtAl2002,StreckerEtAl2002,CornishEtAl2006,MarchantEtAl2013,MedleyEtAl2014,McDonaldEtAl2014,NguyenEtAl2014,MarchantEtAl2016,EverittEtAl2015,LepoutreEtAl2016}
And --- in the presence of an optical lattice --- also for repulsive
interaction~\cite{EiermannEtAl2004}. Recent experimental results for
bright solitary waves in attractively interacting Bose-Einstein condensates include both a first step
to interferometric applications~\cite{McDonaldEtAl2014} and collisions of
two bright solitons~\cite{NguyenEtAl2014} as well as quantum
reflection off a narrow attractive potential
barrier~\cite{MarchantEtAl2016}.

While current experiments can successfully be modeled by using the
mean-field (GPE) approach, beyond-mean-field properties of quantum
bright solitons are a focus of ongoing theoretical
investigations~\cite{CarterEtAl1987,LaiHaus1989,CastinHerzog2001,CarrBrand2004,WeissCastin2009,SachaEtAl2009,StreltsovEtAl2009,BieniasEtAl2011,StreltsovEtAl2011,BarbieroEtAl2014,StreltsovaStreltsov2014,GertjerenkenKevrekidis2015}.   In the current paper, we
explore beyond mean-field many-particle quantum
properties of matter-wave
bright solitons~\cite{GertjerenkenEtAl2012,Gertjerenken2013}
(cf.~\cite{StreltsovaStreltsov2014}) for interferometric
purposes. While for
mean-field based interferometry, quantum fluctuations in, for example,
the center of mass velocity can endanger the interferometric
scheme~\cite{HelmEtAl2014}, in the present paper both interactions and quantum
fluctuations are the tools to generate quantum-enhanced
interferometry. While spin-squeezed states~\cite{KorbiczEtAl2005,Schnabel2008,LiEtAl2008,FerriniEtAl2008,EsteveEtAl2008,GilEtAl2014} are one possibility to
achieve~\cite{GrossEtAl2010} quantum-enhanced interferometry,
Schr\"odinger-cat states\footnote{Schr\"odinger-cat states are
  discussed in
  Refs.~\cite{MonroeEtAl1996,DunninghamBurnett2001,DunninghamEtAl2005,WeissCastin2009,
    StreltsovEtAl2009,GarciaMarchEtAl2011,YaoEtAl2012,DellAnna2012,Gertjerenken2013,FogartyEtAl2013,FischerKang2015}
  and references therein.} are another~\cite{GiovannettiEtAl2004,GiovannettiEtAl2011,ChwedenczukEtAl2011}
(cf.~\cite{BraunsteinCaves1994}). An experimental
  realization of  an atom interferometer can be found in
  Ref.~\cite{DimopoulosEtAl2007}; interferometry using non-entangled
  states leading to enhanced precision can be found in
  Ref.~\cite{BoixoEtAl2008}.  

We use the rigorously proved~\cite{WeissCastin2012}
effective potential approach developed independently of each other in
Refs.~\cite{SachaEtAl2009,WeissCastin2009}. The effective potential
approach is valid in the regime of low kinetic energies where the
soliton is energetically forbidden to classically break apart, thus
corresponding to a mean-field regime where reflection at a barrier
leads to steps in the transmission coefficient~\cite{GertjerenkenEtAl2012,WangEtAl2012,DamgaardHansenEtAl2012}.\footnote{In
  the regime of higher kinetic energy, where mean-field bright
  solitons do break apart when hitting a barrier, scattering has been
investigated, for example, in Refs.~\cite{CuevasEtAl2013,HelmEtAl2015,DunjkoOlshanii2015} and references therein.} In
this energy regime both mesoscopic Schr\"odinger-cat states~\cite{WeissCastin2009,GertjerenkenEtAl2012} and, for
two-component Bose condensates, mesoscopic Bell states~\cite{GertjerenkenEtAl2013} have been
predicted theoretically to occur in the many-particle quantum dynamics
of quantum matter-wave bright solitons.  For a single-species,
attractively interacting Bose condensate and repulsive barrier, the
effective potential approach~\cite{SachaEtAl2009,WeissCastin2009} is
a powerful tool to model the many-particle quantum dynamics both for
solitons of the order of $N=100$ atoms~\cite{WeissCastin2009} but also for
much smaller particle numbers of $N=2$~\cite{GertjerenkenWeiss2012}.

The paper is organized as follows: Section~\ref{sec:modeling}
describes how quantum bright solitons in one-dimensional wave guides can be modeled; {followed by the effective potential approach (Sec.~\ref{sec:effective})} we use to describe the
many-particle quantum dynamics in the regime where matter-wave bright
solitons are energetically forbidden to break apart. In
Sec.~\ref{sec:experimentally} we introduce the signature we propose
to use for future experiments: the probability to find all particles
on the side opposite to the initial condition after the soliton hits
the barrier for the second time. In Sec.~\ref{sec:quantum} we present
our results: the quantum-enhancement achieved by using quantum bright
solitons rather than single particles (or non-interacting
Bose-Einstein condensates). The paper ends with conclusions and outlook
in Sec.~\ref{sec:conclusion}.


\section{\label{sec:modeling}Modeling quantum bright solitons}
In order to model attractively interacting atoms ($g_{\rm 1D}<0$) in
one dimension, the integrable Lieb-Liniger-(McGuire) Hamiltonian~\cite{LiebLiniger1963,McGuire1964} 
\begin{equation}
\hat{H} = -\sum_{j=1}^N\frac{\hbar^2}{2m}\frac{\partial^2}{\partial
  x_j^2}+\sum_{j=1}^{N-1}\sum_{n=j+1}^{N}g_{\rm 1D}\delta(x_j-x_n)
\label{eq:LL}
\end{equation}
is available, where
 $x_j$ denotes the position of particle $j$ of mass $m$.
The (attractive) interaction
\begin{align}
\label{eq:g1d}
g_{\rm 1D} &=2\hbar
\omega_{\perp}a \\ &<0 \nonumber
\end{align}
 is proportional to the
\textit{s}-wave scattering
length $a$ and the perpendicular angular trapping-frequency,
$\omega_{\perp}$~\cite{Olshanii1998}. Including the center-of-mass momentum $K$, 
the (internal) Lieb-Liniger ground state~\cite{McGuire1964,CastinHerzog2001}
reads (cf.~\cite{CastinHerzog2001})
\begin{equation}
\label{eq:wave}
\Psi(x_1,x_2,\ldots,x_N) \propto e^{iKX} \exp\left(-\frac{m |g_{\rm 1D}|}{2\hbar^2}\sum_{j<\nu}|x_j-x_{\nu}|\right);
\end{equation}
the center-of-mass coordinate is given by 
\begin{equation}
\label{eq:CoMCoord}
X = \frac 1N \sum_{j=1}^N x_j.
\end{equation}
The ground-state energy is given by
\begin{equation}
E_0(N,g_{1\rm D}) = -\frac{mg_{1\rm D}^2N(N^2-1)}{24{\hbar^2}}.
\end{equation}
Even more relevant for what we propose here, there is an energy gap
between the internal ground state and the first excited state,
\begin{align}
E_{\rm gap} &\equiv E_0(N-1) - E_0(N)\\
&= \frac{mg_{1\rm D}^2N(N-1)}{8{\hbar^2}}.
\label{eq:Egap}
\end{align}
As long as the total kinetic energy of the matter-wave bright soliton
lies well within this energy gap,
\begin{equation}
  \label{eq:valid}
E_{\rm kin} < E_{\rm gap},
\end{equation}
 the validity of the effective potential approach of
Refs.~\cite{WeissCastin2009,SachaEtAl2009} can be proved
rigorously~\cite{WeissCastin2012}. This parameter-regime coincides
with a complete break-down of the validity of the Gross-Pitaevskii
equation\footnote{\label{footnote:GPE}The
  mean-field approach via the Gross-Pitaevskii equation (GPE)~\cite{PethickSmith2008}
\[
\label{eq:GPE}
i\hbar \frac{\partial}{\partial t}\varphi = -\frac{\hbar^2}{2m}\frac{\partial^2}{\partial x^2}
\varphi
+(N-1) g_{1 \rm D}|\varphi|^2 \varphi.
\]
is a useful tool to explain ongoing state-of-the art experiments with
matter-wave bright solitons like the experiments of
Refs.~\cite{NguyenEtAl2014,MarchantEtAl2016}. However, the mean-field
approach fails in the quantum
regime~\cite{GertjerenkenEtAl2012,GertjerenkenEtAl2013} that makes the
quantum-enhanced
interferometry proposed in the current paper possible.
} for describing scattering matter-wave solitons off, for example, a
narrow barrier~\cite{GertjerenkenEtAl2012,GertjerenkenEtAl2013,WangEtAl2012,DamgaardHansenEtAl2012}.

 If
the center-of-mass wave function is a delta function and the particle
number is $N\gg 1$, then the single-particle density can be
shown~\cite{CalogeroDegasperis1975,CastinHerzog2001} to be equivalent
to the mean-field result based on the Gross-Pitaevskii
equation~\cite{PethickSmith2008} (cf.~footnote~\ref{footnote:GPE})
\begin{equation}
\label{eq:singlesoliton}
\varrho(x)= \frac{N}{4\xi_N\left\{\cosh[x/(2\xi_N)]\right\}^2},
\end{equation}
 normalized to the total number of particles $N$; 
the soliton length is given by
\begin{equation}
\label{eq:solitonlength}
 \xi_N \equiv \frac{\hbar^2}{m\left|g_{\rm 1D}\right|(N-1)}.
\end{equation}

Adding a longitudinal 
  harmonic trapping potential,
\begin{align}
  V(x_1, x_2,\ldots, x_N) &= \sum_{j=1}^N V_0(x_j),\\
V_0(x) &= \frac 12 m\omega^2(x+X_0)^2,
\end{align} 
does not change the physics (beyond
breaking integrability) - as long as~\cite{Castin2009} the soliton length is small
compared to the harmonic oscillator length
\begin{equation}
\lambda_{\rm HO} \equiv \sqrt{\frac{\hbar}{m\omega}},
\end{equation}
The plane waves  in the center-of-mass coordinate in
Eq.~(\ref{eq:wave}) become harmonic oscillator eigenfunctions in this
limit~\cite{HoldawayEtAl2012}. After preparing the system in its
ground state, at $t=0$ the single-particle potential $V_0$ is changed
quasi-instantaneously to
\begin{equation}
\label{eq:V1}
V_1(x) = \frac 12 m\omega^2x^2 + v_1\delta(x-X_{\rm S}),
\end{equation}
thus moving the center of the longitudinal trap by $X_0$
  and adding a very narrow 
scattering potential (modeled by a delta function) which is shifted by a distance of 
$X_{\rm  S}$ from the trap center.  As the proof of the validity of the
effective potential approach introduced in the following section uses repulsive
potentials~\cite{WeissCastin2009,WeissCastin2012} we choose $v_1>0$,
where $v_1$  quantifies the strength of
the scattering potential.

\section{\label{sec:effective}The rigorously proved~\cite{WeissCastin2012} effective potential approach}
{The effective potential method was developed independently by K.~Sacha, C.~A.~M\"uller, D.~Delande, and J.~Zakrzewski~\cite{SachaEtAl2009} and C.~Weiss and Y.~Castin~\cite{WeissCastin2009}. The rigorous mathematical proof by Y. Castin published a couple of years later~\cite{WeissCastin2012} includes strict error bounds demonstrating that this is a highly reliable many-particle method in the energy regime where bright solitons cannot break apart for energetic reasons [Eq.~(\ref{eq:valid})].

While this restricts the applicability to kinetic energies that lie within the energy gap between the internal ground state (all $N$ particles in one soliton) and the internal first excited state ($N-1$ particles in one soliton), the effective potential is a high-end many-particle method that replaces the many-particle Schr\"odinger equation by an effective equation for the center-of-mass wave function in a non- perturbative way~\cite{WeissCastin2012}. In cases where the center-of-mass wave function is localized at two places separated by more than the soliton length, this automatically corresponds to highly entangled Schr\"odinger-cat states.

In general, the effective potential is the convolution between
the scattering potential and the soliton~\cite{WeissCastin2009,SachaEtAl2009,WeissCastin2012}. For the relevant single-particle potential~(\ref{eq:V1}) the result is particularly
simple: the harmonic trap simply is replaced by a
harmonic trap with identical trapping frequency but for a particle of
mass $mN$; the delta function is replaced by the
mean-field soliton 
profile~(\ref{eq:singlesoliton}) multiplied by the
strength of the delta-function scattering potential~$v_1$~(cf.~\cite{WeissCastin2012}). As both potentials are a function of the center-of-mass coordinate, this leads to a further huge advantage of the effective potential approach:
 by introducing the
effective potential}, we have replaced the many-particle Schr\"odinger
equation by an effective single-particle Schr\"odinger equation for
the center of mass~$X$ with  
\begin{align}
\label{eq:SchrEffective}
i\hbar \frac{\partial}{\partial t}\varphi(X,t) =&
-\frac{\hbar^2}{2mN}\frac{\partial^2}{\partial X^2}\varphi(X,t)\nonumber\\
&+V_{\rm eff}(X) \varphi(X,t), \quad E_{\rm kin} < E_{\rm gap},\\
V_{\rm eff}(X) =&  \frac 12
mN\omega^2 X^2
+ \frac{v_1N}{4\xi_N\left\{\cosh[(X-X_{\rm S})/(2\xi_N)]\right\}^2}.
\label{eq:Veff}
\end{align}
 The proof of the effective-potential approach requires
  that the kinetic energy is smaller than the energy gap~(\ref{eq:Egap})~\cite{WeissCastin2009,WeissCastin2012}.

For sufficiently small ratio of soliton length to harmonic oscillator
length, the effective potential can even be
approximated by~\cite{GertjerenkenEtAl2012}
\begin{equation}
\label{eq:canevenbe}
V_{\rm eff}(X) \simeq \frac 12
mN\omega^2 X^2
+ v_1 N\delta(X-X_{\rm S}),
\end{equation}
 which could be treated
  analytically~\cite{BuschEtAl1998}. In the following section, our
  approach will be to combine
an approximate analytical treatment of the scattering of the soliton
off the potential with full 
 numerics for which the center-of-mass wave function of the ground
  state soliton is prepared in the harmonic oscillator ground
  state. The trap is then shifted such that the soliton moves towards
  the new minimum --- in which then a narrow scattering potential is
  switched on.

           While the model displayed in
Eqs.~(\ref{eq:SchrEffective})-(\ref{eq:canevenbe}) looks like a
single-particle approach, these equations for a particle of mass $Nm$
describe many-particle quantum dynamics in the parameter regime in
which quantum bright solitons cannot break apart~\cite{WeissCastin2009,WeissCastin2012} (cf.\
\cite{SachaEtAl2009}). Any solution of Eq.~(\ref{eq:SchrEffective})
which spreads over more than the soliton width can thus be identified
as being a mesoscopic quantum superposition~\cite{WeissCastin2009} of
the Schr\"odinger-cat type relevant for quantum-enhanced
interferometry~\cite{GiovannettiEtAl2004}. In the following, we will
focus on 50:50 splitting leading to the particularly useful~\cite{GiovannettiEtAl2004} high-fidelity Schr\"odinger cat states. To
identify more general quantum superpositions see, e.g., Ref.~\cite{BachRzazewski2004}.

{
To summarize, thanks to the rigorous proof of Ref.~\cite{WeissCastin2012}, the effective potential approach is a highly reliable many-particle method for quantum bright solitons. Its level of reliability in the regime of low kinetic energies (in which the proof of Ref.~\cite{WeissCastin2012} is valid) is at least as high as high-end many-particle methods used, for example, in~\cite{BarbieroEtAl2014,StreltsovEtAl2009} (cf.\ Refs.~\cite{MajorEtAl2014,CosmeEtAl2016b}). The occurrence of highly entangled Schr\"odinger-cat states relevant for quantum-enhanced interferometry~\cite{GiovannettiEtAl2004,ChwedenczukEtAl2011} is thus guaranteed --- under conditions that are within the reach of ongoing state-of-the-art experiments~\cite{MedleyEtAl2014,NguyenEtAl2014,MarchantEtAl2016,EverittEtAl2015}.}

\section{\label{sec:experimentally} Experimentally accessible signature}

\begin{figure}
\includegraphics[width=\linewidth]{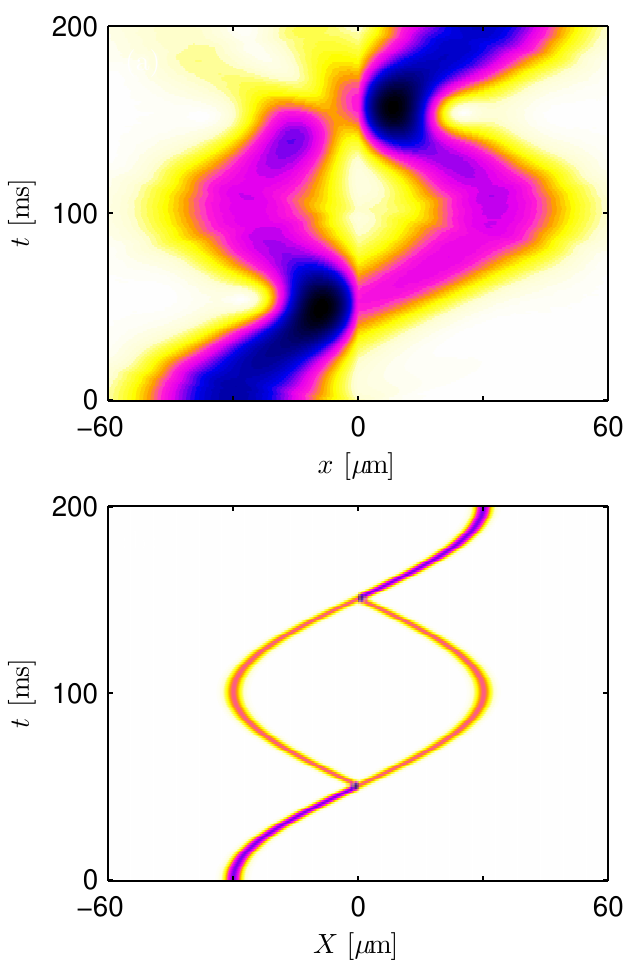}
\caption{\label{fig:signature}(Color online) As the experimentally accessible
  signature we suggest to use the probability to find all particles on
the side opposite to the initial condition after scattering
twice  [shown is the two-dimensional projection of the modulus squared
of the center-of-mass wave function as a function of both position and
time for (a) $N=1$ and (b) a $N=100$ soliton for the parameters of footnote~\ref{footnote:parameters}]. After being prepared in the ground state of a harmonic trap, at
$t=0$ the trap is shifted and a narrow
\textit{repulsive}\/ 
 scattering potential switched
on at the new center of the trap. After the scattering potential is
hit for the first time, Schr\"odinger-cat states occur, after the
second collision all particles are on the side opposite to the initial
condition (cf.\ \cite{GertjerenkenEtAl2012}). The fact that
Schr\"odinger-cat states occur is a consequence of the fact that slow
bright solitons cannot energetically break apart
(cf.~\cite{WeissCastin2009},  footnote~\ref{footnote:parameters}); the
interaction remains switched on the entire time. }
\end{figure}

We propose to prepare the ground state of $N$ attractively interacting
atoms in a one-dimensional wave-guide with additional week harmonic
trapping. After quasi-instantaneously shifting the harmonic potential by several soliton
lengths, we switch on a very narrow
\textit{repulsive}\/ 
 scattering potential in the center
of the trap and observe the quantum dynamics for 50:50 splitting after
the first collision for two collisions (cf.~\cite{GertjerenkenEtAl2012,Gertjerenken2013,FogartyEtAl2013}).
In Fig.~\ref{fig:signature} we compare what happens to a single atom ($N=1$)
to a  $N=100$ quantum bright soliton: In the limit that the effective
potential behaves similar to a delta-function, the length-scales are a
factor of 10 smaller for the $N=100$ soliton\footnote{\label{footnote:parameters}For \textsuperscript{7}Li and $N\approx 100$, the set of
  parameters used is
 the slightly modified parameter set (we doubled the strength of the
 radial
 trapping frequency) of Ref.~\cite{WeissCastin2009} for the \textit{s}-wave scattering
  length~$a= -1.72 \times 10^{-9}\,\rm m$, $\omega_{\perp}=2\pi\times
  2\times 4800\,\rm Hz$. 
In addition we chose $\omega = 2\pi\times 5 \,\rm Hz$. The initial
distance from the trap center is chosen to be
$X_0=30\,\rm\mu m$; this can easily be reduced for bright
solitons (but not for single particles or weakly interacting
Bose-Einstein condensates). For the soliton it
corresponds to $E_{\rm kin}/|E_0(N)|\simeq 0.83$, within the regime of
validity of the effective potential approach [cf.~Eq.(\ref{eq:canevenbe})].
} compared to $N=1$. In
passing we note that the signature demonstrated  in
Fig.~\ref{fig:signature} b (and also in \ref{fig:quantum-enhanced} c) is not accessible on the
mean-field (GPE) level as it involves intermediate
Schr\"odinger-cat states\footnote{\label{foonote:NotGPE}In the energy regime in which the
  many-particle quantum mechanics predicts Schr\"odinger-cat states,
  the GPE shows strong
  jumps~\cite{GertjerenkenEtAl2012,WangEtAl2012,DamgaardHansenEtAl2012}
  in the transmission behavior. When colliding with the scattering
  potential twice, the
  solitons are either reflected or transmitted in both cases. Thus,
  the GPE-soliton ends on
  the same side as the initial condition (cf.~\cite{CuevasEtAl2013}) and on the opposite side to the
  many-particle quantum prediction.} 
 as the soliton is not energetically allowed to
break into smaller parts, thus what is shown in
Fig.~\ref{fig:signature}~b for times $t\approx 100\,\rm ms$ corresponds to a quantum superposition of
all particles being either on the left or right of the scattering
potential. The main difference between the two panels of
Fig.~\ref{fig:signature} is that the upper panel is only for one
particle. While one might choose to call this a Schr\"odinger cat~\cite{MonroeEtAl1996},
if describing a non-interacting Bose-Einstein condensate with $N$
particles in a rather classical product state, the following section
shows that the state displayed in the upper panel lacks the
many-particle entanglement properties necessary for quantum-enhanced
interferometry for which the state displayed in the lower panel is
particularly useful.

In order for the signature displayed in Fig.~\ref{fig:signature}  
to work, the time-scale for decoherence events via,
for example, single- or three- particle losses has to be 
 large
compared to the total time of the experiment from the first time the
soliton hits the barrier to being detected at the side opposite to its
initial condition. Reference~\cite{WeissCastin2009} shows a possible
parameter set for \textsuperscript{7}Li, using
conservative estimates for the loss rates (cf.~\cite{ShotanEtAl2014}),
thus justifying that using trapping frequencies in the direction of
motion of the order of $\omega=2\pi\times10\rm\, Hz$ is indeed justified.

Before we show how the signature displayed in Fig.~\ref{fig:signature}
can be used for future experiments with quantum-enhanced measurements,
we would like to stress again that the fact that we have Schr\"odinger
cat states relies on strong attractive interaction preventing the
soliton from breaking apart and thus allowing the rigorously proved~\cite{WeissCastin2012} effective potential
approach of Refs.~\cite{WeissCastin2009,SachaEtAl2009} Eqs.~(\ref{eq:SchrEffective})-(\ref{eq:canevenbe}) to be used.
While often interaction is only present to generate entanglement, for the signature  displayed in Fig.~\ref{fig:signature} to work it is
furthermore essential to leave the interaction switched on the entire
time --- otherwise the feature that all particles end up opposite to
their initial condition would not work.

\section{\label{sec:quantum}Quantum-enhanced measurement:
  beyond-classical precision}

\subsection{Overview of Sec.~\ref{sec:quantum}}
In this section, we use the mathematically rigorous effective potential approach (Sec.~\ref{sec:effective}) to show that we can obtain beyond-classical precision in interferometric measurements with quantum-bright solitons.  The fact that we are in the regime of validity of the effective potential approach guarantees both the occurrence of Schr\"odinger cat states and ensures that this is not accessible to non-linear physics (for the GPE in the regime of low kinetic energies, the soliton would either pass the barrier twice or be reflected twice, thus in both cases ending back in the original position rather than opposite to the initial position as shown for the quantum case in Fig.~\ref{fig:signature}.

In Sec.~\ref{sub:dependence} we show that the signature to find all
particles opposite to the initial condition oscillates as a function
of the distance of barrier from the center of the trap (moved by a
horizontal force) position [Eq.~(\ref{eq:T})] both by using an
intuitive analytical approximation which is backed up by systematic
numerical simulations. In Sec.~\ref{sec:catmeasure} we
  summarize the role  Schr\"odinger cats play in our scheme and
  stress the role of measuring atoms in Bose-Einstein condensates via
  scattering or absorbing photons. Section~\ref{sub:precision} shows how this is related to the precision: as for the quantum-enhanced interferometry with Schr\"odinger-cat states  described in Refs.~\cite{GiovannettiEtAl2004,GiovannettiEtAl2011} gain a factor of $N$ in the precision with which the phase can be detected. Repeating the single-particle experiment $N$ times more often than the experiment with $N$-particle Schr\"odinger-cat states yields an overall win of a factor of $1/\sqrt{N}$ for the Schr\"odinger cats~\cite[Sec.~``Quantum-Enhanced Parameter Estimation'']{GiovannettiEtAl2004}. In Sec.~\ref{sub:gravity} we show how well horizontal differences in the gravitational field could be measured.

\subsection{\label{sub:dependence}Dependence of the signature on the distance of the barrier from the trap center}
\begin{figure}[h!]
\includegraphics[width=0.9\linewidth]{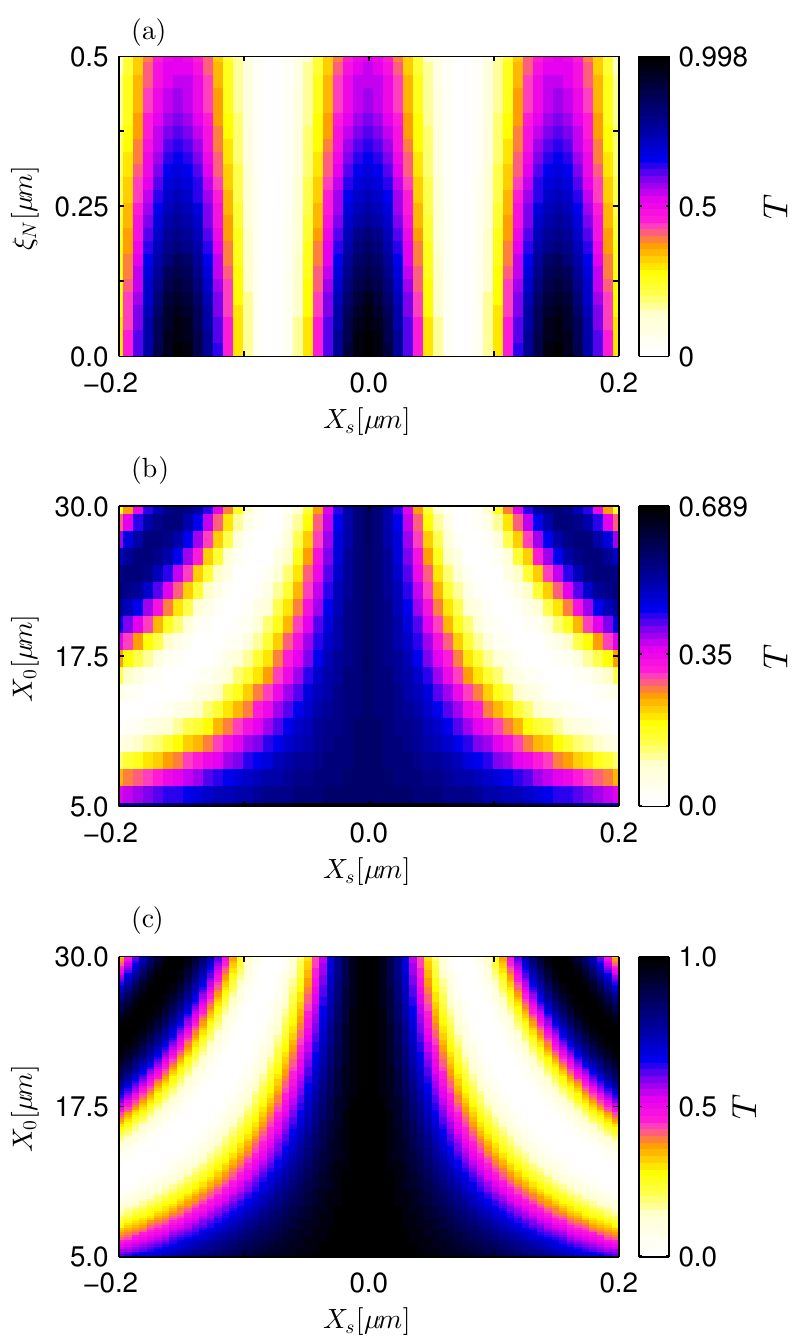}
\caption[??]{\label{fig:systematic} (Color online) 
(a) Our signature that everything is on the side opposite to the
initial condition after colliding with the barrier twice, denoted by
$T$, as a function of distance $X_{\rm{s}}$ between trap minimum and
scattering potential for different widths of the
$1/\cosh^2$-potential, introduced in Eq.~(\ref{eq:Veff}),
$X_0=30$~$\mu$m and $N=100$ particles. The value $\xi = 0.42$~$\mu$m
corresponds to the parameters from footnote
\ref{footnote:parameters}. The limit $\xi \to 0$ corresponds to a
delta function potential. (b) Same for varying initial displacements
$X_0$ and $\xi = 0.42$~$\mu$m kept fixed. (c) Displays the
analytical result~(\ref{eq:T}) for the same parameters as panel
(b). Thus panel (c) confirms that the analytical
approximation~(\ref{eq:T})  qualitatively correctly
capture the physics displayed in the many-particle quantum dynamics of
panel (b).
}
\end{figure}

As a practical example showing how much this improves the precision, we
perform a thought measurement of
a horizontal gradient in the gravitational potential, modeled as a
linear potential added to Eq.~(\ref{eq:SchrEffective}),
\begin{equation}
V_{\rm gravity}(X) =- XmN \Delta g_{\rm gravity},
\end{equation}
which would lead to a shift of the zero of the harmonic trapping potential and thus to a
non-vanishing distance
\begin{equation}
X_{\rm S} = \frac{\Delta g_{\rm gravity}}{\omega^2}
\end{equation}
 between trap center and scattering potential. In order to estimate
 the precision of our interferometric measurement, we need to determine
 the position of a maximum by repeating the measurements $n$ times.
The interference pattern as a function of $X_{\rm S}$ --- namely the
transmission coefficient after scattering at the potential twice ---
is given by the approximate analytical formula~\cite{GertjerenkenEtAl2012} 
\begin{align}
\label{eq:T}
  T &= \frac 12 \left[1+\cos(4\Omega X_{\rm s})\right]\\
  \Omega &= \frac{Nm\omega X_0}{\hbar}.
\label{eq:Omega}
\end{align}
In order to derive the last line, we used energy conservation to
derive the center-of-mass momentum $\hbar K$ via 
\begin{align}
  \frac{(\hbar K)^2}{2Nm} &= \frac 12 N m \omega^2 X_0^2,\quad {\rm
    and~thus}\\
K &= \frac{Nm\omega X_0}{\hbar},
\end{align}
together with the text-book~\cite{Fluegge1990} result that scattering
a particle of mass $Nm$ with a plane wave with momentum $\hbar K$ from
a delta function potential
\begin{equation}
V(X) = \frac{\hbar^2\Omega}{Nm}\delta(X)
\end{equation}
leads to fifty-fifty splitting if
\begin{equation}
K = \Omega.
\end{equation}
While the true effective potential will be broader than the
delta-function used here, this only affects the amplitude, not the
spatial period of the oscillations (see Fig.~\ref{fig:systematic}~a,
cf.~\cite[Fig.~3]{GertjerenkenEtAl2012}). For the same parameters as
the numerical data depicted in Fig.~\ref{fig:systematic}~b,
Fig.~\ref{fig:systematic}~c shows the analytical formula~(\ref{eq:T})
confirming that this formula predicts the width of the interference
patterns correctly.  As predicted by
Eq.~(\ref{eq:Omega}), the spatial period increases for decreasing
$X_0$ (see Fig.~\ref{fig:systematic}~b). Thus the delta-function
approximation is sufficient for the purpose of estimating the precision
of our interferometric setup.

\subsection{\label{sec:catmeasure}Role of Schr\"odinger-cat states \&
  measuring particles in a BEC}

The signature we use for our interferometry scheme relies on a very
fundamental level on Schr\"odinger-cat states: after two collisions
with a barrier, neither the numerical
results for the transmission after two collisions with a barrier depicted in Fig.~\ref{fig:systematic} nor the analytic result
of Eq.~(\ref{eq:T}) could have been obtained with a mean-field
description via the GPE. In the regime of low kinetics energies in
which the rigorously derived effective potential approach
is valid (see Sec.~\ref{sec:effective},
\cite{SachaEtAl2009,WeissCastin2009,WeissCastin2012}), the GPE predicts
unphysical jumps in the transmission/reflection behavior~\cite{GertjerenkenEtAl2012,WangEtAl2012,DamgaardHansenEtAl2012}; solitons
would be either reflected twice or transmitted twice, in both
cases ending up at the same side of the scattering potential for the
two collisions investigated here, corresponding to the mean-field
(GPE) prediction
\begin{equation}
T_{\rm GPE} = 0.
\end{equation}
Rewriting Eq.~(\ref{eq:T}) to show the $N$-dependence explicitly, 
\begin{equation}
T =  \frac 12 \left[1+\cos\left(N \frac{4m\omega X_0 X_{\rm S}}{\hbar}\right)\right],
\end{equation}
yields a reduction of the width of the interference patterns by a factor of
$N$. This is identical to the general case discussed in
Ref.~\cite[Sec.~``Quantum-Enhanced Parameter Estimation'']{GiovannettiEtAl2004} when replacing a single-particle
quantum superposition by an $N$-particle Schr\"odinger cat, the first step towards
quantum-enhanced interferometry.

Observations of atomic Bose-Einstein condensates employ scattering or absorption of laser light~\cite{StamperKurnKetterle2001}.
Multiple scattering let alone absorption of the same
photon are not an issue that is discussed in such
cases~\cite{StamperKurnKetterle2001}, a fact that is further stressed
by the small atom numbers $N\approx 100$ which are the focus of our
current paper.
 Thus, measuring a BEC of $N$ noninteracting
atoms is equivalent to measuring $N$ single atoms independently. In
both cases, the central limit theorem thus increases the precision by a
factor of $1/\sqrt{N}$.

Note that if measuring the Schr\"odinger cat state, an individual
measurement of $T$ is either 0 or $N$. While having $N$ atoms in a
bright soliton can simplify detecting it, there is no additional correction
corresponding to the factor  $1/\sqrt{N}$.

To determine the precision, in Sec.~\ref{sub:precision} repeat the above thought experiment $n$
times (comparing an non-interacting BEC with the bright soliton
Schr\"odinger-cat state).

\subsection{\label{sub:precision}Precision}
 
In order to answer how precisely we can determine the position of the
scattering potential with respect to the trap center, we have to
consider a couple of separate points. All are based on the distance between
minima/maxima, as determined by Eq.~(\ref{eq:T}). Half the
distance between two minima is given by
\begin{align}
\Delta X_{\rm S} &= \frac{\pi}{4\Omega}\\
&= \frac 1N \frac{\pi\hbar}{2m\omega X_0}.
\label{eq:precisionNref}
\end{align}

\begin{figure}[h!]
\includegraphics[width=0.9\linewidth]{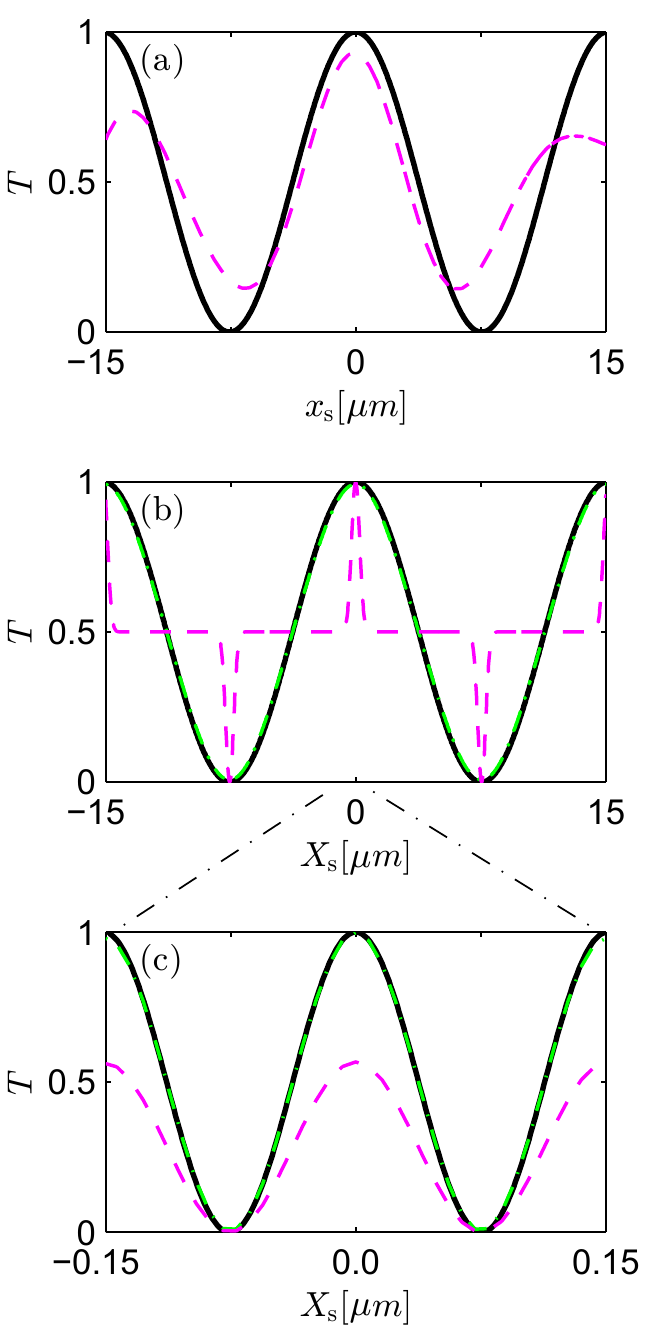}
\caption[???]{\label{fig:quantum-enhanced}(Color online) Our signature that everything
  is on the side opposite to the initial condition after colliding
  with the barrier twice as a function of distance between trap
  minimum and scattering potential [see.~Fig.~\ref{fig:signature} also
  for parameters used] for
  (a) a single particle: numerical data [magenta (dark gray) dashed curve] and
  analytical formula~(\ref{eq:T}) (black solid curve) and (b) repeating the experiment 100 times with
  single particles [magenta (dark gray) dashed curve] compared to the single-particle
  result~(\ref{eq:T})  (black solid curve) and replacing this result
  by a combination of Gaussians  [green (light gray)
  dash-dotted curve]. (c) Shows quantum-enhancement by a factor of
  $1/\sqrt{N}$  for a quantum bright soliton of
  $N=100$. Shown are formula~(\ref{eq:T}) (black solid curve),
  numerical data using a delta-function barrier [green (light gray)
  dash-dotted curves] and the $1/\cosh^2$ potential corresponding to the parameters  of
  footnote~\ref{footnote:parameters} [magenta (dark gray) dashed curve].}
\end{figure}

\begin{enumerate}
\item The aim is to move the scattering potential back to the center when it was moved by a small distance from there; thus our device will have to remember by how much the potential has moved.
\item Repeating the experiment $n$ times for various distances will give us a part of the cosine of Eq.~(\ref{eq:T}) centered around the maximum.
\item Half the distance between two minima gives a first order of
  magnitude of the precision with which we can measure the position;
  the larger the particle number in our soliton the higher the
  precision will be. (In an actual experiment,  the full width at half
  maximum could be used equivalently).
\item\label{point:made} If measured in terms of the precision that can be read off the
  distance of the interference fringes, a quantum-enhanced measurement
  is better by a factor of $1/N$. This is the typical scaling for
  quantum-enhanced interferometers with Schr\"odinger-cat states as
  compared to single particles~\cite[Fig.~1]{GiovannettiEtAl2011}. This
  difference of a factor of $N$ is visible by comparing
  Fig.~\ref{fig:quantum-enhanced}~a and
  Fig.~\ref{fig:quantum-enhanced}~c.
\item In our case, the increased precision described in
  point~\ref{point:made} above is related to the de Broglie wavelength
  of the particles; smaller wavelengths allowing to measure smaller
  distances. Thus, this is independent of precise details of the
  initial state.

\item Repeating the quantum-enhanced measurement $n$ times with
  Schr\"odinger-cat states of $N$ atoms in order to
  determine the position of the maximum involves $n\times N$ atoms.
\item In order to compare the outcome with the single-particle case
  (cf.~\cite{GiovannettiEtAl2011}) thus repeat the same interferometric
  sequence of measuring the 
  position of the maximum with  single particles $n\times N$
  times. Compared to the quantum-enhanced case, this increases the
  number of measurements by a factor of $N$, allowing us to determine
  the position of the maximum more precisely. As the errors involved
  in repeating such experiments $N$ times more often scale as 
  $1/{\sqrt{N}}$~\cite[Sec.~``Quantum-Enhanced Parameter Estimation'']{GiovannettiEtAl2004}, 
\begin{equation}
\delta X_{N~\rm non-interacting~particles}\propto \frac{\delta X_{N=1~\rm non-interacting~particle}}{\sqrt{N}}
\label{eq:noninter}
\end{equation}
 we thus have
\begin{align}
\delta X_{N-{\rm particle~soliton}} \propto \frac{\delta X_{N~\rm non-interacting~particles}}{\sqrt{N}} \\
\propto  \frac{\delta X_{N=1~\rm non-interacting~particle}}{{N}}
\label{eq:inter}
\end{align}
\end{enumerate}

Thus, to summarize the precision of the single-particle case is limited by the Heisenberg-uncertainty relation [the distance measured~(\ref{eq:precisionNref}) is of the order of the de Broglie wavelength; independent of the precise initial state, the momentum is of the order of $m\omega X_0$]. The minimal distance measurable can be improved by statistical analysis of repeating the experiment $N$ times, giving an improvement of a factor of  $1/\sqrt{N}$ for the smallest distance that can be measured [Eq.~(\ref{eq:noninter})] for repeated experiments with single particles~\cite[Sec.~``Quantum-Enhanced Parameter Estimation'']{GiovannettiEtAl2004}. A precision which scales faster with $N$ as displayed in Eq.~(\ref{eq:inter}) thus seems, at first glance, to violate the Heisenberg uncertainty relation.

However, many-particle entangled states like Schr\"odinger-cat states are known to show such behavior which is sometimes described as ``beating the standard quantum limit''~\cite{GiovannettiEtAl2004}. As the de Broglie wavelengths are a factor of $N$ smaller than for the single-particle case, it should not be suprising that we can measure distances with the precision of Eq.~(\ref{eq:inter}). The overall gain of a factor of $1/\sqrt{N}$ is of course consistent with the Heisenberg-uncertainty relation if applied to our $N$-particle Schr\"odinger-cat state (cf.~\cite{GiovannettiEtAl2004,GiovannettiEtAl2011}).

While the analytical calculations are based on the approximate result
given in Eq.~(\ref{eq:T}), the oscillation periods are confirmed by
doing full numerics (Figs.~\ref{fig:systematic} and~\ref{fig:quantum-enhanced}): In panel
Fig.~\ref{fig:quantum-enhanced}~a, the full numerical result shows
that it is straightforward to identify the position of the
maximum. Combining the central-limit theorem with the analytical
approximation using Gaussians demonstrates that repeating the
experiment 100 times will lead to an increased precision for the
position of the maximum of a factor of 10 (Fig.~\ref{fig:quantum-enhanced}~b).
While the numerics using the delta-function agrees
  with the analytic predictions of  Eq.~(\ref{eq:T}), for the
  $1/\cosh^2$ potential corresponding to the parameters  of
  footnote~\ref{footnote:parameters} the
  amplitude is smaller.

\subsection{\label{sub:gravity}Measuring horizontal differences in the gravitational potential}

Thus, our thought
experiment can measure differences in the gravitational acceleration
of at least
\begin{equation}
|\Delta g_{\rm gravity}| < \omega^2\Delta X_{\rm S}
\end{equation}
which yields a difference in acceleration of
\begin{equation}
|\Delta g_{\rm gravity}| <  \frac 1N \frac{\pi\omega\hbar}{2m X_0}.
\end{equation}
This yields a difference in the acceleration potential of 
\begin{align}
\label{eq:precision}
2 X_0|\Delta g_{\rm gravity}| <  \frac 1N \frac{\pi\omega\hbar}{m},
\end{align}
which takes into account the fact that we effectively
  average $\Delta g_{\rm gravity}$ over a distance $2 X_0$. Thus,
  within the definition~(\ref{eq:precision}), the precision is
    independent of our choice of $X_0$. The strength of quantum
  bright solitons lies in particular in the fact that one can get much
  closer to the scattering potential than for non-interacting
  particles, thus exploring shorter length-scales.

For \textsuperscript{7}Li, $N=100$ (cf.~\cite{WeissCastin2009}, footnote~\ref{footnote:parameters}) and
$\omega = 2\pi\times 5\,\rm Hz$, if our thought experiment involves
measuring $n$ solitons, it can detect
differences smaller than $(0.009/\sqrt{n})\,\rm\mu m\, m/s^2$.
 By replacing
\textsuperscript{7}Li by \textsuperscript{85}Rb we could gain
another order of magnitude in precision.
 While at first glance it might appear to be tempting to propose to both further reduce
  the trapping frequency and further increase the particle number, the
  validity of Eq.~(\ref{eq:precision}) relies on the occurrence of
  intermediate Schr\"odinger-cat states and thus will work better for
  not too large particle numbers and not too long time-scales.

Depending on the magnitude of the potential differences we intend to
measure on micrometer scale  (gravity was
just an example), the precision might actually be too large to easily
determine the distance of the scattering potential from the trap
center. To work out a clever way to start with reduced precision
(with non-interacting atoms and/or higher trapping frequencies) in
order to roughly determine the position and to
continuously increase the precision could still be an engineering
challenge. Nevertheless, our results clearly indicate the feasibility of such a device.

\section{\label{sec:conclusion}Conclusion and outlook}

Similar to the case of the spin-squeezed states realized in the experiment of
Ref.~\cite{GrossEtAl2010}  demonstrating interferometric precision surpassing
what is possible classically, Schr\"odinger-cat states can also be
used for quantum-enhanced measurements~\cite{GiovannettiEtAl2004,GiovannettiEtAl2011} (cf.~\cite{ChwedenczukEtAl2011}). Quantum bright solitons generated from attractively interacting Bose gases are an ideal system to produce such Schr\"odinger-cat states~\cite{WeissCastin2009}; state-of-the art experimental setups with ultra-cold atoms~\cite{MedleyEtAl2014,NguyenEtAl2014,MarchantEtAl2016,EverittEtAl2015} could reach such a regime in the near future.  Ultracold atoms can be applied
 to questions not easily accessible to  photon experiments (cf.~\cite{ScullyDowling1993}).

In the present paper, we propose to use small, harmonically trapped quantum matter-wave
bright solitons for quantum-enhanced measurements. The signature we
propose to measure --- the probability to find all particles at the
side opposite to the initial condition --- is not accessible on the
mean-field (GPE) level as it involves intermediate
Schr\"odinger-cat states (see
footnote~\ref{foonote:NotGPE}).

 Our
numerical simulations are done using the effective potential approach
of Refs~\cite{WeissCastin2009,SachaEtAl2009} by taking care that the
prerequisite of the proof~\cite{WeissCastin2012} is fulfilled: The
solitons are energetically forbidden to break apart. Thus, if the
center-of-mass wave function is split 50:50 when the soliton hits the
barrier for the first time, we are certain to have Schr\"odinger cat
states. For the second step to work (that after recombining at the
barrier the soliton ends at the side opposite to the initial condition,
it is furthermore essential to maintain the quantum
superposition~\cite{GertjerenkenEtAl2012}. While experimentally
observing the motion of quantum bright solitons in the presence of
decoherence via atom losses~\cite{WeissEtAl2015} would also be an
interesting experiment, the quantum-enhanced interferometry will only
work if decoherence via particle losses happens on time-scales longer
than the duration of the experiment (cf.~\cite{WeissCastin2009}).

 When shifting the position of the scattering
potential away from the center, our signature oscillates. The spatial
oscillation period scales $\propto 1/\sqrt{N}$ with the number of
particles, thus providing the ``usual'' increase in precision by a factor of
$1/\sqrt{N}$ quantum-enhanced measurements can provide. Here, it
occurs when the position of the maximum is determined via repeated measurements.

A potential experimental realization of such a
setup might be useful
to measure small horizontal gradients in the gravitational force on
micrometer scales  in an approach complementary to existing experiments combining
Bose-Einstein condensates and gravity~\cite{MuentingaEtAl2013}. 

The data presented in this paper
are available online~\cite{GertjerenkenEtAl2016Data}.

\acknowledgments

We thank
 T.\ P.\ Billam, 
S.\ L.\  Cornish, S.\ A.\ Gardiner, S.\ A.\ Hopkins, R.\ G.\  Hulet, and  L.\ Khaykovich for
discussions.
B.~G.\ thanks the European Union for funding through FP7-PEOPLE-2013-IRSES Grant Number 605096.
C.~W.\ thanks the UK Engineering and Physical Sciences Research Council (Grant No.\ EP/L010844/1) for funding. 
The computations were performed on the HPC cluster HERO, located at the
University of Oldenburg and funded by the DFG through its Major Research
Instrumentation Programme (INST 184/108-1 FUGG), and by the Ministry of
Science and Culture (MWK) of the Lower Saxony State.
%

\end{document}